**REVIEW**                                                                                                                   **Open Access**

# IoT trust and reputation: a survey and taxonomy

Muhammad Aaqib, Aftab Ali[*], Liming Chen and Omar Nibouche

## Abstract

IoT is one of the fastest-growing technologies and it is estimated that more than a billion devices would be utilised across the globe by the end of 2030. To maximise the capability of these connected entities, trust and reputation among IoT entities is essential. Several trust management models have been proposed in the IoT environment; however, these schemes have not fully addressed the IoT devices' features, such as device's role, device type and its dynamic behavior in a smart environment. As a result, traditional trust and reputation models are insufficient to tackle these characteristics and uncertainty risks while connecting nodes to the network. Whilst continuous study has been carried out and various articles suggest promising solutions in constrained environments, research on trust and reputation is still at its infancy. In this paper, we carry out a comprehensive literature review on state-of-the-art research on the trust and reputation of IoT devices and systems. Specifically, we first propose a new structure, namely a new taxonomy, to organise the trust and reputation models based on the ways trust is managed. The proposed taxonomy comprises of traditional trust management-based systems and artificial intelligence-based systems, and combine both the classes which encourage the existing schemes to adapt these emerging concepts. This collaboration between the conventional mathematical and the advanced ML models result in design schemes that are more robust and efficient. Then we drill down to compare and analyse the methods and applications of these systems based on community-accepted performance metrics,e.g. scalability, delay, cooperativeness and efficiency. Finally, built upon the findings of the analysis, we identify and discuss open research issues and challenges, and further speculate and point out future research directions.

**Keywords** Trust management, Reputation, Taxonomy, Internet of Things, Data Mining, Artificial intelligence

## Introduction

Internet of Things (IoT) is one of the fastest emerging technologies and it is estimated that approximately 50 billion devices, including everything from smart phones to home appliances, will be used around the world by the end of 2030 [1]. IoT refers to the network of various physical objects connected through the internet that can exchange data and information [2, 3]. Due to innovation in sensor technology, IoT is used in multiple applications like home automation, defense system, self-driven cars, monitoring of the environmental system, industrial IoT (IIoT), and retail shops, etc. [4, 5]. However, the primary focus of trust and reputation in IoT devices is on the trust across IoT layers' architecture, applications, and devices. The development of a smart environment is greatly aided by the enhancement of user privacy and authentication, both of which are made possible through the application of trust and reputation in data communication. According to Gambetta [6], trust is a specific level of individual personal judgment by which an agent expects another agent to undertake a specific type of action on which the agent's security and authentication welfare is dependent. One possible method for calculating trust is reputation-based, which bases trust on reputation metrics gathered and connected across distributed environments. Therefore, an entity's Reputation (RP) is an expectation of its behaviour based on observations made by other entities about the entity's previous acts in a certain context at a given moment [7].

*Correspondence:
Aftab Ali
a.ali@ulster.ac.uk
School of Computing, Ulster University, Northern Ireland BT15 1ED, UK

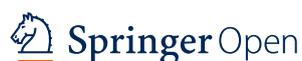

Because of fluctuations in wireless channels, mobility, and lack of battery power, IoT devices are differentiated by extremely dynamic node connections and network topology. These aspects of the IoT are what lead to a node disconnecting from the network. IoT networks present new security risks that might have serious consequences for an organisation's finances and reputation. In most essential applications, including remote healthcare, vehicle communications, etc., these threats might have fatal consequences, including inaccessible resources, unauthorised access, information disclosure, and data and resource damage. Several trusts and reputation frameworks [8–10] have been proposed as a solution to this issue of device security for safeguarding the devices from being attacked and damaged i.e., unavailability of resources. Many traditional trust and reputation techniques are based on the previous knowledge of the relationship between the trustee and a trustor, and when there is no prior knowledge then an appropriate evaluation of trust and reputation cannot be established. Similarly, trust calculation performance metrics suffer greatly when insufficient data are employed and the calculation technique prohibits meaningful data information from being gathered. However, in many traditional trust calculation systems, trust is determined by weighing and calculating the trust factors. Moreover, to ensure the accuracy of an evaluation method, it is complicated to conclude the weights. Therefore, due to these types of challenges, there is a need to develop a robust scheme for IoT trust and reputation calculations. In our classification, methods are categorised into two classes (i.e., traditional trust management-based, and Artificial intelligence-based schemes) depending on the performance metrics used for the IoT trust and reputation evaluation. These two classes cover both traditional and AI-based trust management schemes (TMS), as well as present a systematic survey of state-of-the-art TMS. In this paper, we put forward an in-depth survey of TMS for IoT devices. The key contributions of this study are:

- A comprehensive analysis of the published research on IoT TMS in the scientific community.
- Introducing an improved taxonomy of the trust and reputation of IoT devices.
- A detailed analysis of the design approaches, key performance metrics for evaluating the efficiency and accuracy of traditional as well as artificial intelligence-based frameworks, advantages, and disadvantages of the most important trust and reputation management schemes.
- Identification of a number of open issues and challenges for promoting research on trust and reputation in the realm of IoT devices.

The remainder of the paper is structured as follows. The existing literature on the topic undertaken is presented in the section titled 'Existing survey'. Section 'Methodology' describes the methodology and research selection based on inclusion and exclusion criteria. Section 'Classification and taxonomy of trust models' overviews the taxonomies of existing schemes. Section 'Trust and reputation evaluation metrics and useful terms of IoT' illustrates several performance metrics for assessing trust and reputation in IoT devices. Section 'Artificial intelligence based models' presents the discussion on existing artificial intelligence-based models. Section 'Open issues and challenges' interprets the open problem and challenges. Lastly, the section 'Conclusion' concludes our work.

**Existing surveys**

This section presents existing surveys in the field of trust and reputation using IoT devices. Sicari et al. [11] identified numerous open research problems in the heterogeneous environment of the IoT. However, in a heterogeneous environment, meeting security requirements such as data privacy, authentication, and trust between users and things is critical. A study in [12] highlighted the importance of information related to trust and reputation in providing an environment where reliable and trustworthy data is required for IoT users. The authors discussed five methods for obtaining reliable data for IoT devices. Eder et al. [13] carried out a review of existing trust and reputation-based frameworks in the Internet of Things domain. They discussed methods for gathering information, grading, and transaction approval. A comprehensive examination of previous techniques to manage trustworthy activities together with the features of TMS for IoT is highlighted in [14]. The author proposed the objectives of IoT trust management and designed a holistic TMS to specify the research challenges. Guo et al. [15] is a review of several trust evaluation schemes from the perspective of identity management in IoT devices. Moreover, they identified the already established trust evaluation system based on the gathering of information and trust aggregation. The authors presented an analysis of frequently and rarely visited trust evaluation models after summarising the pros and cons of each. Harwood and Garry [16] proposed an analytical understanding of trust, reputation, and its possible promising applications. They provide practitioners with a perspective that helps them to develop an appropriate TMS in the context of specific techno-service scenarios. Wang et al. [17] suggested a review of trust calculation schemes, mentioning a specific standard of trustworthiness, availability, and reliability. The author highlights and focuses on some research gaps for the calculations of trust and contributes to future investigation in the IoT domain. Nguyen et al. [18] analysed several trust evaluation schemes like centralised and decentralised systems for the identification of several future developments. Furthermore, the author did a comparative analysis and determined the characteristics of trust, performance metrics, and advantages and disadvantages. Suryani et al. [19] examined a number of TMS, taking into account their vulnerabilities to various threats and trust features. However, the work lacks a comparative analysis of IoT trust models. Din et al. [20] surveyed TMS aimed for IoT devices. Based on a thorough investigation of trust management, relevant strategies are classified, with benefits and limitations listed. The authors also looked at the advantages and disadvantages of various TMS for IoT devices. Behrouz et al. [21] carried out a review of existing schemes in the IoT TM. Their scheme were categorised based on 4 classes namely, reputation-based, policy-based, prediction based and recommendation based. Then they reviewed and compared employing several trust metrics. Additionally, the authors discussed the benefits and drawbacks of the schemes in each class. The reviews and advancement of the previous work related to TM in IoT were carried out by the authors [22]. They analysed a number of challenges in IoT devices that are raised in terms of management, heterogeneity, integrity, and scalability, as well as thoroughly discussed TM issues from various aspects. The authors of [23, 24] investigated TMS designed for the IoT. They examined and compared various designs used to develop TMS for the IoT. The characteristics of such schemes are then examined to demonstrate the similarities and differences between existing IoT TMS. Additionally, a comparative analysis was carried out to group the key characteristics that contribute to both efficiency and resilience. The TM schemes in [25] are used to categorise distributed and centralised trust models. A centralised TMS typically assumes the third party guarantees the privacy of the data. An efficient and reliable data management system for IoT is provided by a reliable central server, which also facilitates system implementation.

The aforementioned surveys focus on the taxonomy of traditional trust and reputation evaluation techniques, as well as trust and related attack characteristics. Furthermore, existing works fail to address a broader and more adequate scope for trust and reputation. Similarly to [19], the authors focused on comparing the trust models in terms of various trust properties and lacks the classification of trust models. Similarly, the authors in [26, 27], perform a wider study of trust and reputation, including centralised and decentralised systems, dynamic behavior, and future challenges, but did not mention the precise taxonomy for trust and reputation. There is no all-encompassing management platform for combining a multi-trust taxonomy. Therefore, there is still a need for a comprehensive review of trust and reputation schemes using artificial intelligence mechanisms along with traditional models. Most existing surveys [11, 14–16, 22] discuss improving accuracy, scalability, and flexibility, as well as highlighting the characteristics of trust management in IoT devices, but they do not focus on integrity, availability, and privacy. Furthermore, some of the works [11, 12, 15, 16, 19, 21] highlighted the importance of TMS in IoT devices and classified the taxonomy based on traditional techniques. However, most schemes struggle to identify the most contributing parameters for calculating trust and reputation. The main aim of the proposed survey is to evaluate and classify the computation and evaluation schemes of trust and reputation for IoT devices. The current work gives a comprehensive survey based on 2 classes, i.e., traditional and artificial intelligence-based TMS. These two main classes are more appropriate than those discussed in the literature because they cover most of the traditional and AI-based TMS. However, existing surveys [12, 15, 16, 19, 25] focus on a single class and are primarily concerned with traditional TMS, leaving out AI-based TMS. Similarly, this review of the trust management and reputation models will lead to identifying consistent factors, parameters, and methods, which will help in the future development of trust and reputation management frameworks. Furthermore, the proposed survey is the first to provide a new taxonomy based on traditional and AI-based models and provide a comparative study of IoT trust and reputation mechanisms.

Table 1 presents a complete summary of the existing surveys and thereby depicts a clear picture of trust and reputation as a security problem in the IoT environment. A broad analysis of IoT trust models, including their focuses and contributions, which were not addressed in prior surveys such as single and multi trust, design taxonomy, are summarized in this table. Moreover, it highlights the limitations in the literature by providing a comprehensive examination of trust and reputation in IoT devices, along with the comparison of traditional methods and AI techniques. The specific ways in which AI techniques can enhance trust and reputation in IoT devices, such as the use of machine learning algorithms for trustworthy and non-trustworthy devices, are also highlighted. Furthermore, the open issues and challenges associated with implementing these methods and techniques are discussed and specific solutions for addressing these issues and challenges are provided. Finally, a detailed overview of the current state-of-the-art trust and reputation in IoT devices, and offers a valuable resource and direction for researchers in TMS is put forth.

**Table 1** Summary of related TMS surveys

| Ref | Multi Trust | Traditional methods | Comparison AI models | Taxonomy classes | research issues | Contributions |
| --- | --- | --- | --- | --- | --- | --- |
| Zheng et al. [14] **2014** | ✗ | ✗ | ✗ | ✗ | ✗ | Analysis of trust characteristics for research direction |
| Wang et al. [17] **,2016** | ✓ | ✗ | ✗ | ✗ | ✗ | Concentrated on comparing trust schemes in terms of trust features |
| Chen et al. [15] **, 2017** | ✓ | ✗ | ✗ | ✓ | ✓ | Discussed stages of TMS |
| Din et al. [20] **, 2018** | ✓ | ✓ | ✗ | ✓ | ✓ | Evaluation of TMS, including pros and cons |
| Behrouz et al. [21] **, 2019** | ✗ | ✓ | ✓ | ✓ | ✗ | Analysis of TMS used in the IoT |
| Abbas et al. [28] **, 2019** | ✓ | ✓ | ✗ | ✓ | ✗ | Classified trust schemes based on IoT functions requirement |
| Mohammad et al. [29] **,2022** | ✓ | ✓ | ✓ | ✗ | ✓ | Analysis of the edge computing |
| Alyzia et al. [30] **, 2022** | ✓ | ✓ | ✗ | ✓ | ✗ | Categorized works based on TMS |
| Ebrahimi et al. [23] **, 2022** | ✓ | ✓ | ✗ | ✗ | ✗ | Simulations and quantitative comparisons are used to determine the factors |
| Our contributions **, 2023** | ✓ | ✓ | ✓ | ✓ | ✓ | Focused on new design approach for classification and key metrics for evaluating the accuracy of TMS |

## Methodology

The proposed system uses a systematic literature review to verify precise, accurate data search and retrieval. Initially, we developed a review protocol to specify the research strategy, inclusion, and exclusion benchmark for selecting the scientific article to be reviewed and choosing an aim to analyse the chosen papers. The main steps included in the proposed approach to the review work are summarised in Fig. 1.

### Research selection methodology

Initially, we conducted a systematic review of those findings that well-defined IoT devices' privacy, security, trust, and reputation calculation models. The whole research was conducted through the most important available databases, consisting of ACM, IEEE-Explore, Elsevier, and Springer. They represent the current state of knowledge because they contain publications from reputed journals and conference proceedings on IoT device security, trust, and reputation. Constraining the research to four intimately available databases suggests that a pattern of the literature in the recommended study is targeted. Moreover, the boundary of the review work was constrained to the domain of computer science (CS), computer science and theory (CST), and engineering studies. In the preliminary stages, the research was made through filtering conferences, technical reviews, and journal articles that were presented and suggested between 2013 to March 2022. The following terms are used to retrieve the manuscripts.

- "Trust" AND "IoT device" OR "Reputation" AND "IoT device" "Trust" OR "Reputation" AND "IoT Device"
- "Trust" AND "TMS" OR "DTMS" OR "IoT Device"
- "Trust" AND "Reputation" OR "Data Mining" OR "Machine/Deep Learning"

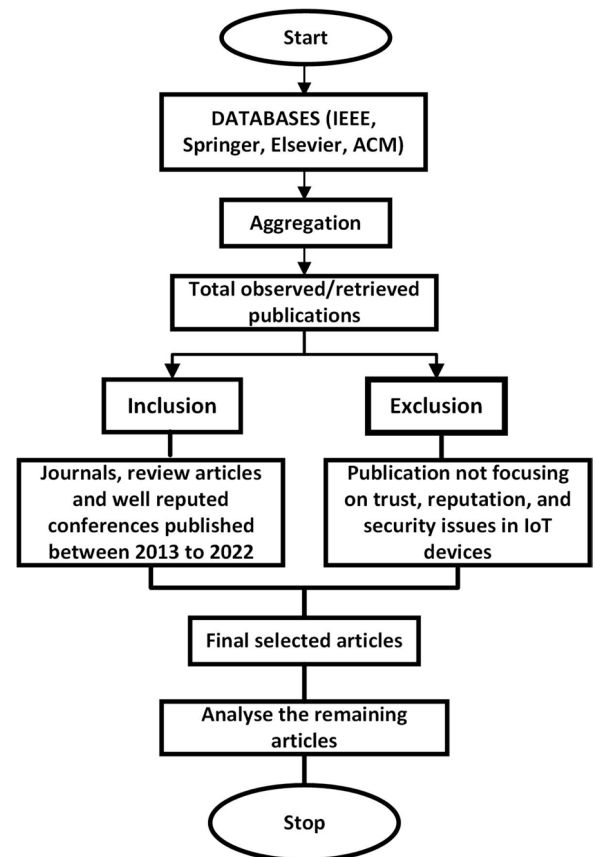

**Fig. 1** Methodology for inclusion and exclusion

**Table 2** Management of inclusion and exclusion

| Applying Inclusion | Applying exclusion |
| --- | --- |
| Published between 2013 - 2022 | Survey and review papers in taxonomy |
| Field of IoT devices security | Not peer-reviewed procedure |
| Trust and reputation | Not completely available on the online database |
| Existing quality aspects of publication | Not focusing on trust, reputation, and security in IoT devices |

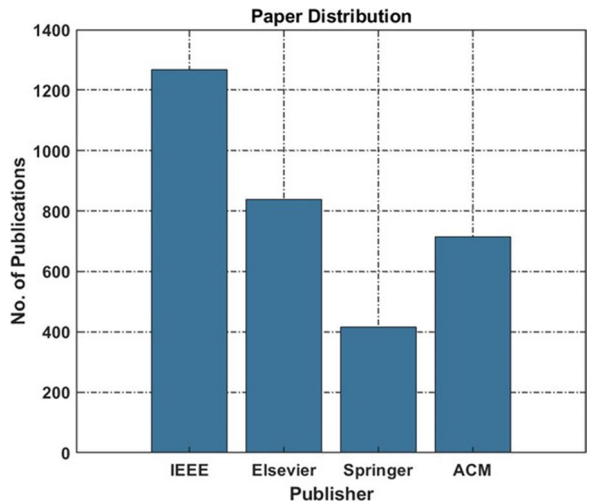

**Fig. 2** Article distributions based on digital databases

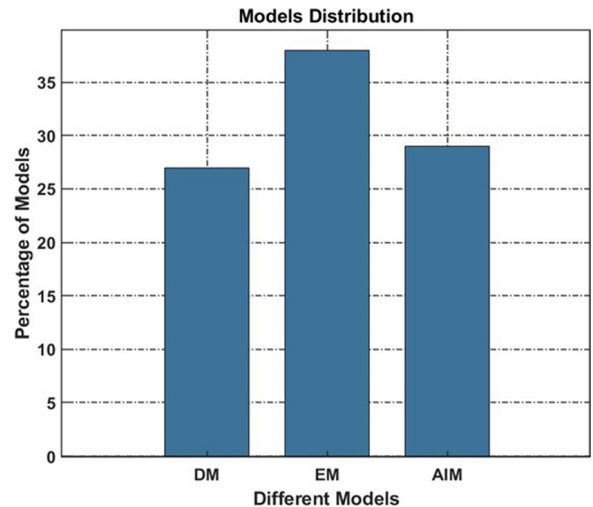

**Fig. 3** Distributions of manuscripts based on traditional and AI-based model

A total of 1267 articles from IEEE Xplore, 417 from Elsevier, 838 from Springer, and 715 from ACM have been obtained. Again, a total of 3237 manuscripts were observed, and a check was done on the title and abstract of the investigated articles. After the scanning process, 3052 manuscripts were noticed that were not relevant to the extent of the proposed review and were eliminated. Then, the criteria of inclusion and exclusion were applied to the remaining 185 scientific articles used for further processing. Furthermore, we highlighted some of the inclusion and exclusion requirements in Table 2 for article consideration. Among 185 articles, 120 fulfilled the inclusion requirements. Lastly, the survey paper distribution according to four digital databases is revealed in the following Fig. 2.

Most of the manuscripts were chosen from the IEEE Journals (i.e.,39%), followed by springer 26%, the next well-established journals are the Association of Computing Machinery (ACM) (i.e., 22%) and finally, 13% taken from the journal of Elsevier respectively. The papers and articles included in the taxonomy have been categorised by decision models (DM), Evaluation models (EM), and Artificial Intelligence based models (AIM) as depicted in Fig. 3.

**Research questions**

It must be possible to formulate specific research questions once the theoretically appropriate research phenomena have been found. The following research questions are focused on TMS in IoT for problem identification and improving work precision:

- RQ1 Which schemes and models are currently employed in IoT for trust calculation?
- RQ2 What type of classification was used for IoT trust and reputation systems?
- RQ3 What characteristics are used to assess the establishment of trust?

We addressed RQ1 by going over each scheme's techniques in considerable details and evaluating the advantages and disadvantages of IoT TMS. In order to present the TMS used, we respond to RQ2 by dividing the survey paper into two major classes. Finally, we addressed RQ3 by comparing several trust features in IoT trust and reputation management systems. Additionally, we have organised these features into the table based on their simulation environment.

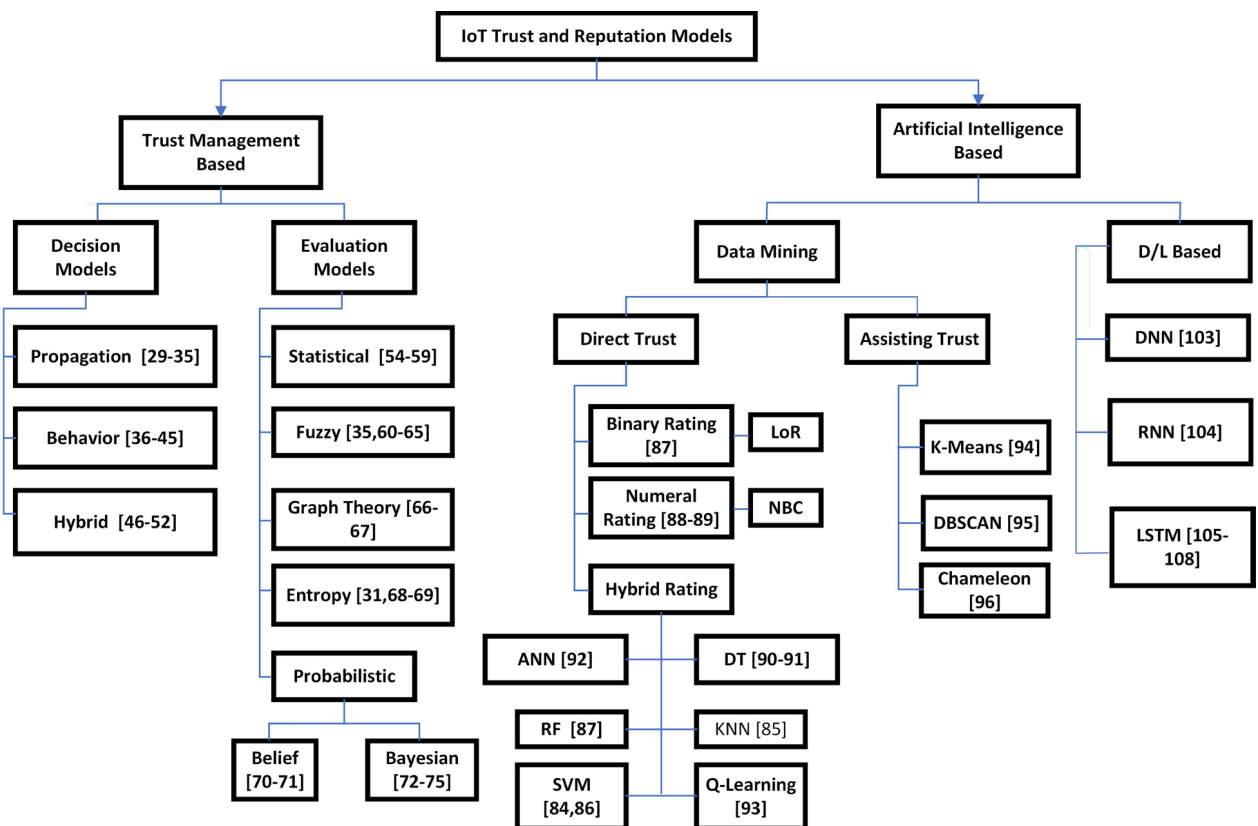

**Fig. 4** Taxonomy of IoT devices trust and reputation

## Classification and taxonomy of trust models

In this section, we develop a thematic taxonomy for the IoT device's trust and reputation. This taxonomy is built based on two main factors (e.g. Trust management based on evaluation and decision models, and artificial intelligence based on data mining and deep learning frameworks) as described in Fig. 4. The first factor described some of the key traditional methods to identify trust and reputation in IoT devices and prevent malicious behavior and attacks. Similarly, the second factor revealed AI-based trust and reputation frameworks to prevent devices from attacks and manage a large amount of data and classify them accordingly. The following sections illustrate each of these methods and their shortcomings in further detail.

### Trust management-based model

Many researchers have proposed a trust-based model for calculating the trustworthiness of IoT devices. Many different formulas for calculating trust scores have been reported by researchers, and some of them have a view of trust which is binary that is trustworthy or untrustworthy.

*Decision models (DM)*

The policy-based technique is employed based on the decision model and is also called policy maker [26]. As a result, the authors have declared the development of a policy service capable of operating in extremely complex relationship environments with entities while also performing well in large-scale environments. For the purpose of a factory supply chain using IIoT, Mohammad et al. [31] proposed TMS to execute a multi-agent scheme and evaluate the efficacy of three different techniques: SIoT, reputation-distribute-conflict (RDC), and Reputation in Gregarious Societies (RegReT).RegReT outperformed the other two techniques in terms of evaluating trustworthiness using simulations. The decision models can be categorised into three sub-sections i.e., propagation, behavior, and hybrid [27].

Propagation: Trust propagation (TP) is a method in which an IoT node generates its trust estimation outcome across another node. However, the TP can be achieved in two ways (i) node point and (ii) cluster point. In node point propagation, the devices propagate trust scores independently without any usage of a cluster head. The authors in [32, 33] have presented a scheme based on the node-level propagation model of trust and reputation. At the cluster level, the cluster head controls all the connections which are made by the inter and intra cluster [34]. In this approach, the authors investigated lightweight trust computing and employed the model of propagation. The authors [35] suggested a trust communications protocol in Social IoT (SIoT). They employed a scheme of trust propagation and aggregation to satisfy the accuracy and resilience of the system. Moreover, they examined the understanding of trust convergence and its variations and achieved a community of interest (CoI) and honesty in a dynamic SIoT environment. In the traditional system, centralised trust propagation (CTP) is connected with the cloud-based trust and reputation (TR) scheme. A centralised TR scheme is implemented employing the structure of a hash table as described in [36]. A system using a centralised TMS was developed by Din et al. [37] for lightweight IoT devices. In order to exchange services between devices, the scheme manages certificates of trust without performing trust calculations. In terms of cooperation and compatibility, they also quantified additional observations of direct trust. On the other hand, trust recommendations are used to observe indirect trust. Researchers in [38] identified the main entities of trust and recommended additional research on the paradigm of trust propagation. However, the major threat to the cloud-based CTP is the pattern of the framework that could assist the information of TP data between cloud servers and IoT devices.

Behavior: Mayano et al. [39, 40] presented a behavior model that explains that the trustor utilises their experiences and observations through the trustee and computes the level of trust. In [41] the authors presented a behavior-based reputation system for the node to establish trust. However, they also propose a trust architecture that merges software-defined networks in IoT and layers of protocol authorisation in trust management to solve the scalability issues in the dynamic environment. Another good example is presented in [42] which suggested a TMS, to evaluate trustworthiness by utilising direct observations. The developed system returns the assisting entities with an authentic score and punishes the malicious node's score negatively. To satisfy the model against on-off attacks (OOA), the authors analysed the system and identified the malicious behavior based on two factors, 1) the position of the nodes and 2) density of malicious nodes. The results exhibited that the performance of the developed model was effective and accurate against OOA in the network system. However, the robustness of the system has not been tested against other types of attacks. The scheme in [43] presented new trust management and retrieval system called predictability trust that differentiates between interim errors and malicious behaviours and predicts the future values of trust based on previous behaviours with an adaptable design.

In [44] a model was developed to control information about the IoT device's reputation. The proposed model is capable to reduce the effects of malicious behaviors that have been considered. The model also provides a defensive mechanism and detects malicious activities. However, the established method cannot calculate the trustworthiness of the devices. Another behavior-based scheme is described in [33] which is based on the process of trust management knowledge and quantitatively compared trust management techniques. They suggested a model of trust and reputation based on behaviours to implement assistance between things in an IoT network. Moreover, the established method defends distributed sensor networks (DSN) in IoT against untrustworthy node attacks. The TMS could encourage associations with distributed computing and communications nodes to provide malicious node identification and aid the decision-making methods of various protocols. Abbas et al. [45] developed a dynamic trust management model (DTMM) where the network nodes evaluate the behavior of the peer nodes autonomously and dynamically present rewards and punishments. The developed method effectively identified the OOA and malicious nodes and categorise them based on three levels such as mild, moderate, and severe. Based on experimentation, the DTMM showed better performance than the state-of-the-art models in terms of requiring a minimum time to effectively detect misbehaving nodes.

The work in [46] is also based on the user behaviour of the trusted model, to detect the anomalous behaviour of the user history pattern. The established model considered several types of indication, including security, authentication, operation, and efficiency, to calculate the past behaviours of the user and compare them with the current state of the user's behavior. Moreover, the developed framework employed fuzzy logic to compute the values of comprehensive and direct trust. Nitti et al. [47] investigated the trustworthy system in SIoT and recommend objective and subjective methodology while dealing with the most entities that are evident during dynamic behavior. Moreover, it was essentially invulnerable to the social network's standard activities, like a non-trustworthy person changing their behaviours based on the interactions. The scheme in [48] presented a trust-based model and measured multiple trust issues associated with the actions of nodes. The most prominent factors for trust measurement are the capacity of packet forwarding, rate of repetition, delay, cooperativeness, and integrity. Moreover, Information entropy theory (IET) was used to eliminate the subjective effect of trust factors. Furthermore, the concept of Dempster shaper (DS) was used to determine and identify trust. The developed algorithm showed that the quantitative model of trust achieves good performance against attacks.

Hybrid: A hybrid trust system incorporates the history and recommendation of trust. History and recommendation-based trust calculations are utilised to decide whether the node is trustworthy or not [49]. Sicari et al. [50] suggested a hybrid network architecture using WSNs and wireless mesh networks and supporting a reliable set of information, allowing the network to maintain data trustworthiness. This model shows a good effect on the reduction of data. However, this analysis does not take into consideration data privacy. Kogias et al. [51] elaborate a hybrid model for SIoT based on the COSMOS network and combined a standard solution for IoT that is utilised in Peer to Peer (P2P) and mobile ad hoc networks. An entity in this model computes the trust index based on direct observations. Furthermore, the entity can deter its reputation through the consultation of friends or the consultation of the COSMOS network and platform. In their simulation, it was determined that the model can eliminate the malicious entity from the platform and can obtain high accuracy and low computational cost. However, to establish the authentic behaviour of the node, a probabilistic investigation should be taken into account for each of the COSMOS platforms. The author in [52] examines the trust node management (TNM) of VANETs, which intends to determine the credibility of the node as part of an evaluation process and prevent sending malicious nodes to networks. Initially, a combined trust consisting of direct and recommended trust is presented for each node, and then the trust evaluation process and data communication call for completeness. Finally, the author introduced a time sliding window and a delay function to confirm that the newest data communication has a greater weight. Another good example of a hybrid trust management model (HTMM) is presented in [53] where the authors elaborated on the main challenges with already established trust management frameworks (TMF). The analysis identified two novel attacks i.e., a misbehavior attack and a location-dependent attack against the existing trust establishment technique. The proposed HTMM carries objective characteristics according to which trust for a node is evaluated based on not only direct interactions but indirect data as well. Narang et al. [54] proposed a 'Hybrid Trust Management Framework (HTMF)' employing a method of probabilistic neighbourhood overlap to estimate a strong tie between entities. The probabilistic concept is adapted from past work on social issues and then applied to directed social networks (DSN). The two types of social networks in the established scheme that are applied are 1) probabilistic concept is applied to a social grid and produces multiple networks such as an online social network (OSN) and an IoT device social network. Furthermore, for trust management, both human and artificial intelligence are used, and 2) Trust management is based on both dynamic and static approaches, so it helps control the resource overhead of a dynamic scheme. The developed method has limited applicability and is considered effective only against attacks on IoT devices. In [55] the authors formulate a hybrid system to evaluate the trustworthiness of the sources. To calculate trust value, cloud services are evaluated based on compliance and reputation. The user's reputation is calculated based on their collective feedback, and their feedback rating is based on their perception of the services invoked. The proposed approach has high efficiency for cloud applications. The existing TMS could be categorised into the reputation and trust-based schemes. However, further, improvement is required in the established models to enhance the credibility, flexibility, and dynamicity of the developed systems in IoT devices.

*Evaluation models (EM)*
The evaluation models (EM) are employed to calculate the trust by employing attributes of a certain entity. The authors [56] labeled each entity with trust in order to compare it to individual trust features that influence trust. Furthermore, the model is classified based on evaluation techniques like statistical, fuzzy, graph theory, probabilistic, belief theory, entropy and Bayesian inference models.

Statistical: The statistical models are employed to calculate the feedback provided by transactions and interactions based on mathematical calculations. Sharma et al. [57] used a simple mathematical technique to calculate the trust score. In a statistical model [58], the authors investigated the fact that each node is accountable for computing the trust values in a distributed system. However, a node has some properties to evaluate trust, which include computation power (CP), context awareness (CA), response, and trust. Moreover, the evaluation of trust can be notified through direct and indirect observations. The autonomous system for self-driving cars is part of the IoT, so a trust management scheme has been developed for the autonomous scheme [59]. The authors included an estimation technique to provide safety from spoofing attacks in the model. The proposed model was expected to make a change between adversarial noises and other disruptions, and then a statistical technique was employed to ensure the safety requirements. To compute the value of trust a reputation and recommendations-based model was presented in [8] to enable entities in SIoT to develop relationships in a trustworthy way.

Recommendation and reputation parameters were used to develop a numerical model to estimate trust in each entity. In addition, they simulate and analyse the scheme's effectiveness in comparison with previous approaches and provide a trustworthy model in a smart environment. Moreover, they examined the main characteristics such as convergence, precision, and resiliency against malicious behaviours through experimentation.

Another framework that computes the trustworthiness of the system, especially for distant verification processes based on IoT architecture is explained in [60]. To calculate the trust values, the security factor is also considered during malicious node detections. A proposed model uses trust metrics based on information collected from an aggregation of data nodes and cluster head nodes. The authors focused on developing a decentralised method for calculating trust levels among nodes in a network. The scheme in [61] is also statistical based that calculates the trustworthiness of the IoT device by considering both the user's trust and stereotypical public reputation. In [62] a semi-centralized trust model (SCTM) was created to manage blockchain across multiple domains. The devices are associated in a centralised way and managed by a server that operates the ledger rating data that assists the exchange of data across domains using the consent protocol. The suggested SCTM aggregates direct and indirect knowledge of trust to calculate the trust values of IoT devices.

Fuzzy models: To handle the level of uncertainty and ensure data efficiency, the fuzzy system was adopted. Chen et al. [33] studied the node's trustworthiness and conclude its future behavior, that is how it will react to future operations. In [38], a scheme of distributed trust management (DTM) technique for IoT was developed. They extract the elements of the trust management model, considering the service and decision-making processes, from the trust-based simulations. Observing the service scheme, they developed a system for IoT sensors, a core, and application layers. Furthermore, the proposed scheme employed fuzzy set theory (FST) to execute the trust mechanism of the IoT's layered architecture. Similarly, the authors of [63] employed a fuzzy technique in a trust-based decision-making control mechanism. To calculate the trust values the scheme employed linguistic variables such as recommendation, knowledge, and experience. These linguistic variables are plotted to get approvals for the IoT network. This approach highlighted the scalability of the system and does not affect productivity while increasing the number of devices. A formal (TMS) based on the design of IoT was discussed in [64]. The IoT model was decomposed into three layers; i) sensor layer (ii) core layer and (iii) application layer. Subsequently, each layer is used by TMS for the purpose of effective routing, self-organised and multi-service. The decision-making process was executed by the service requester based on the gathered information and requester policy. The proposed framework does not mention the validation of the TMS for the heterogeneous environment in IoT. A model presented in [65] specified a trust-aware access mechanism that employed an architectural source model (ASM) compatible defense system for the IoT. In contrast to previous systems that only focused on reputation and response, this multidimensional process considered multiple factors such as safety features, QoS, RP, and social interactions to get an approved decision. Moreover, an implementation of the proposed model for constrained and non-constrained devices was carried out on a real testbed. Furthermore, the developed model does not take into consideration the calculation of accuracy and confidentiality.

In [66], the authors proposed a fuzzy-based trust model in a smart grid environment to identify untrustworthy nodes and evaluated it with a previous system to validate its improvements. However, the existing method has not mentioned the calculation of trust and reputation in IoT devices. The fuzzy logic-based technique presented in [67] to identify untrusted nodes is concerned with biased service provisioning as well as on-off attacks. Moreover, the authors developed a trustworthy messaging approach for communication between nodes to maintain the safety of the IoT system. Through the experimental analysis, the developed model exhibited effectiveness under various conditions, such as the detection of on-off attacks and a secured messaging scheme for IoT nodes. However, the model lacks scalability, energy efficiency, and data storage for IoT communication. Similarly, a framework for the allocation of spatial tasks based on an employee's perceived level of trustworthiness is calculated in [68]. the developed method calculates the employee's trust and reputation based on previous interactions and the history of sentiment analysis. Moreover, the authors achieved the values of trust and reputation and employed a technique of fuzzy inference system (FIS) to get the degree of legitimacy of each employee for multidimensional tasks. There are many research challenges in a fuzzy-based scheme such as lack of trust aggregation scheme, and scalable IoT context. Therefore, further improvements are required to achieve the trust service domain in the IoT. As a first step, mechanisms could be developed to manage reputations that motivate entities to publish feedback securely and eliminate trust-related attacks. The second step could be the establishment of intelligent FIS for dealing with trust management knowledge that prefers the best approach in real-time while adapting autonomously changing the environment. However, trustors need to have a chance to express their personal preferences during the TM and trust aggregation (TA) evaluation and calculation processes. The disadvantages of fuzzy-based trust evaluation models are the development of complicated procedures and the setting of membership functions.

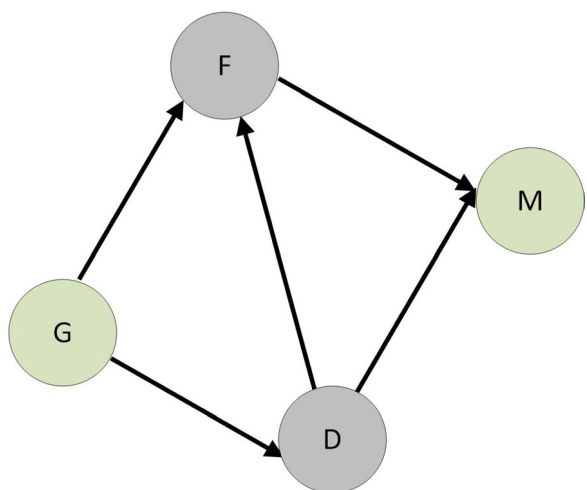

**Fig. 5** The trusted graph for the graph theory scheme

Graph theory: The graph theory-based trust model shows the trust levels of nodes that have opinions about each other through various sequential and parallel paths as shown in Fig. 5. The trust-based graph will implemented when there are a lot of iterative recommendations and observations obtained from the neighbour nodes [69]. The authors presented a trust evaluation scheme based on the Multiplicative Attribute Graph Model (MAGM) in [70]. The developed method calculates the robustness of the trust among social IoT members based on a pre-defined set of social interactions.

Entropy: The authors in [34] proposed a trust management framework based on multiple source information fusion for IoT devices. The proposed scheme used a multiple source information scheme for trust evaluation, followed by a trust evaluation scheme for IoT device cooperativeness and low overhead trust computation schemes. Finally, the authors adopt the fusion scheme based on entropy to overcome the drawbacks of the existing approach. The authors [71] presented a trust assessment framework for a specific IoT domain that carries out trust assessment and computes the trustworthiness level of an IoT device before allowing it to be included in the space. In addition to evaluating a specific device's trustworthiness, the developed challenge-response trust evaluation system gives a trustworthy source that can be applied to direct trust and recommendation trust. Shuzhi Nie [72] proposed a dynamic trust model based on the entropy approach, with the model based on the degree of trustworthy entities and nodes classified into different classes. The created classes and entropy weight schemes are evaluated using the entropy weight method and the degree of a trustworthy node is updated on a regular basis. The method fails to detect several internal attacks because it just identifies the malicious nodes in the network.

Probabilistic models: These models used different schemes and probability distribution functions to calculate the trust score which is further categorised in two ways:

Belief theory: This algorithm is used with other systems like probability and imprecise theories for reasoning with uncertainty. Feng et al. [73] proposed a novel algorithm of belief theory of trust evaluation to integrate the techniques of evidence theory and nodes behavioral scheme. Anwar et al. [74] suggested a trust calculation scheme based on belief theory that identifies non-trustworthy nodes from trustworthy nodes and prevents the system from several attacks. The Bayesian assessment technique was employed to gather direct and indirect values of trust from sensor nodes, and the data was correlated over time and imprecise knowledge was assessed for secure data delivery by avoiding malicious nodes. The proposed approach just mentions attacks and does not alter the trust and reputation calculation of IoT devices.

Bayesian inference: This model allows the probability distribution in case of new observations to deal with various trust variables. According to [75], Bayesian inference with the model parameter of trust to calculate average trust based on a beta reputation scheme. Agus et al. [76] build a model of trust called Trust-Bayes to show a level of trust in IoT. Lin et al. [77] presented a trust evaluation system based on a Bayesian Network (BN). To determine the trust level and calculations of trustworthy nodes the author mentioned the degree of trustworthy nodes and the distribution of conditional probability, respectively. In their experimentation, it was concluded that the efficiency and subjectivity of the evaluation technique could be minimised to some degree. However, this method has limited applicability when varying the distribution of a constant state. Singh et al. [78] suggested a TMS, and its robustness was designed based on 'Bayesian learning and Collaboration Filtering'. To quickly update the behavioural changes, the developed TMS is updated consistently after a specific period with a significant value of decay to the current computed scores. However, the Bayesian inference has some drawbacks when calculating trust, such as complicating the subjectivity of trust with the randomness possibility.

This section highlights the research gap for the aforementioned traditional techniques of trust and reputation in the IoT. However, it is quite evident from the literature that the existing models of security threats and confidentiality must be extended to incorporate trust and reputation using artificial intelligence for the future of IoT infrastructure.

## Trust and reputation evaluation metrics and useful terms of IoT

The concept of IoT is to link everything at any time, anywhere utilising any accessible resource. The goal is to increase the enterprise value of objects by generating interactions between multiple entities. Such diverse relationship requires methods to aid 'things' communicate irrespective of their variety, scalability, energy, recommendation, and dynamicity to establish quite trustworthy and connected IoT devices. Therefore, a TMS is required to tackle these issues in IoT devices. In the following section, we explore the literature and compare the performance metrics for trust and reputation in IoT. As shown in Table 3, the already established systems presented various performance metrics for calculating trust and reputation in IoT devices.

Accuracy is a statistical measure that represents the proximity between standard results and a true value.

Context-awareness (CA) is a fundamental property of trust and should support the evaluation system for the trust and reputation. For example, if the application environment or context changes, then the evaluation system should respond to it and adaptively arrange itself to be able to fit into a different circumstance. According to [102], "Context-aware systems provide related information and services to users depending on their tasks". Context-aware systems use information from their surroundings, such as time, location, and social attributes, to predict the system's demands and adapt their behaviour accordingly.

**Table 3** Comparative Analysis of Trust-Based Models

| Ref. | TR | REP | EN | SC | DE | SEC | CoP | SN | DY | CA | CO | Com | Acc | RC |
|---|---|---|---|---|---|---|---|---|---|---|---|---|---|---|
| Saied et al. [79] **2013** | ✓ | ✓ | ✗ | ✓ | ✗ | ✗ | ✗ | ✗ | ✓ | ✓ | ✓ | ✗ | ✓ | ✓ |
| Wang et al. [38], **2013** | ✓ | ✗ | ✗ | ✗ | ✗ | ✓ | ✓ | ✗ | ✓ | ✗ | ✗ | ✓ | ✓ | ✗ |
| Roman et al. [80], **2013** | ✓ | ✗ | ✗ | ✓ | ✓ | ✓ | ✗ | ✗ | ✗ | ✗ | ✗ | ✗ | ✓ | ✗ |
| Nitti et al. [47], **2014** | ✓ | ✓ | ✗ | ✓ | ✓ | ✗ | ✗ | ✓ | ✓ | ✗ | ✓ | ✗ | ✗ | ✓ |
| Tormo et al. [81], **2015** | ✗ | ✓ | ✓ | ✓ | ✗ | ✗ | ✗ | ✗ | ✓ | ✗ | ✓ | ✗ | ✓ | ✗ |
| Khan et al. [82], **2016** | ✓ | ✓ | ✗ | ✓ | ✗ | ✓ | ✓ | ✗ | ✗ | ✗ | ✗ | ✓ | ✗ | ✗ |
| Chen et al. [83], **2016** | ✓ | ✓ | ✓ | ✗ | ✓ | ✗ | ✓ | ✓ | ✗ | ✗ | ✗ | ✗ | ✓ | ✓ |
| Kogias et al. [51], **2016** | ✓ | ✓ | ✗ | ✓ | ✗ | ✗ | ✓ | ✗ | ✓ | ✓ | ✓ | ✓ | ✓ | ✓ |
| Troung et al. [84], **2016** | ✓ | ✓ | ✗ | ✗ | ✗ | ✓ | ✓ | ✓ | ✗ | ✗ | ✓ | ✗ | ✗ | ✓ |
| Asiri et al. [85], **2016** | ✓ | ✓ | ✓ | ✓ | ✗ | ✓ | ✗ | ✗ | ✓ | ✗ | ✓ | ✗ | ✗ | ✓ |
| Emad et al. [58], **2016** | ✓ | ✗ | ✗ | ✓ | ✗ | ✓ | ✗ | ✗ | ✗ | ✗ | ✓ | ✗ | ✗ | ✗ |
| De Meo et al. [44], **2017** | ✓ | ✓ | ✗ | ✗ | ✗ | ✓ | ✗ | ✓ | ✗ | ✗ | ✗ | ✗ | ✓ | ✓ |
| Coa et al. [86], **2017** | ✓ | ✗ | ✗ | ✗ | ✓ | ✓ | ✗ | ✗ | ✗ | ✗ | ✓ | ✗ | ✓ | ✗ |
| Yu et al. [48], **2017** | ✓ | ✗ | ✓ | ✓ | ✓ | ✓ | ✓ | ✗ | ✗ | ✗ | ✗ | ✓ | ✓ | ✓ |
| Xu et al. [87], **2018** | ✓ | ✗ | ✓ | ✓ | ✗ | ✓ | ✗ | ✗ | ✓ | ✗ | ✗ | ✗ | ✗ | ✗ |
| Dahman et al. [88], **2018** | ✓ | ✗ | ✗ | ✗ | ✗ | ✗ | ✗ | ✗ | ✗ | ✗ | ✗ | ✗ | ✗ | ✓ |
| Zhu et al. [89], **2018** | ✓ | ✗ | ✗ | ✗ | ✓ | ✗ | ✗ | ✗ | ✗ | ✗ | ✗ | ✗ | ✗ | ✗ |
| Azad et al. [90], **2018** | ✓ | ✓ | ✓ | ✗ | ✗ | ✗ | ✓ | ✗ | ✓ | ✓ | ✓ | ✗ | ✓ | ✗ |
| Mahmud et al. [91], **2018** | ✓ | ✗ | ✓ | ✗ | ✓ | ✗ | ✗ | ✗ | ✗ | ✗ | ✗ | ✓ | ✓ | ✗ |
| Malik et al. [92], **2019** | ✓ | ✓ | ✗ | ✗ | ✗ | ✓ | ✓ | ✗ | ✓ | ✓ | ✗ | ✓ | ✗ | ✗ |
| Sheth et al. [93], **2019** | ✗ | ✗ | ✓ | ✓ | ✓ | ✓ | ✗ | ✗ | ✗ | ✗ | ✗ | ✗ | ✗ | ✗ |
| Dorri et al. [94], **2019** | ✓ | ✗ | ✓ | ✓ | ✓ | ✓ | ✓ | ✗ | ✗ | ✗ | ✗ | ✓ | ✗ | ✗ |
| Ursino et al. [95], **2020** | ✓ | ✓ | ✓ | ✗ | ✗ | ✓ | ✗ | ✗ | ✗ | ✓ | ✓ | ✗ | ✗ | ✗ |
| Qiu et al. [96], **2020** | ✓ | ✗ | ✓ | ✗ | ✗ | ✓ | ✗ | ✗ | ✗ | ✗ | ✗ | ✗ | ✗ | ✗ |
| Bo et al. [97], **2020** | ✓ | ✗ | ✗ | ✗ | ✗ | ✓ | ✓ | ✗ | ✓ | ✗ | ✗ | ✗ | ✗ | ✓ |
| Abbasi et al. [98], **2021** | ✓ | ✗ | ✗ | ✗ | ✓ | ✓ | ✗ | ✓ | ✓ | ✗ | ✗ | ✗ | ✗ | ✗ |
| Lingda et al. [99], **2021** | ✓ | ✓ | ✗ | ✓ | ✓ | ✓ | ✗ | ✗ | ✓ | ✗ | ✗ | ✗ | ✓ | ✗ |
| Cheng et al. [100], **2021** | ✓ | ✗ | ✓ | ✓ | ✓ | ✓ | ✓ | ✗ | ✓ | ✓ | ✓ | ✓ | ✗ | ✗ |
| Ikram et al. [37], **2021** | ✓ | ✗ | ✗ | ✓ | ✗ | ✓ | ✗ | ✓ | ✗ | ✓ | ✗ | ✓ | ✗ | ✗ |
| Yaseen et al. [101], **2021** | ✗ | ✓ | ✗ | ✗ | ✗ | ✓ | ✓ | ✗ | ✗ | ✗ | ✗ | ✓ | ✓ | ✗ |

Computational overhead (CO): The efficiency of TMS is determined by the CO analysis, faster the algorithm run, will lower the CO.

Delay: The time period required to transmit a packet from source to target, and it depends on the congestion in the network, as a result, it decreases the network QoS.

Recommendation: Recommendation-based trust is related to specifically recommending a node among other applicable nodes. There are several recommendations and the assigned values to every recommendation are based on trustworthy or untrustworthy transactions. The decision is taken based on the values of the requested recommendation.

Dynamicity: The TMS should be capable to run under conditions, where the services and nodes are accessible or inaccessible at an unpredictable period. The dynamicity makes TMS capable of adapting intelligently to such dynamic situations.

Communication: When there is a small interval between the source and targeted nodes, then the communication trust would depend on direct packet transfer. If the number of communicating packets and relationships between nodes is limited to reflect on trustworthiness, then an optimal choice would be to rely on recommendations of common neighbors between source and destination.

Social network (SN): SN is an important parameter in SIoT for trust evaluation and computation because nodes in a general or standard environment are assumed to have a closer link and to provide the value of trustworthiness.

Cooperativeness: It is the degree of social cooperation between the trustee and the trustor. According to [35] the cooperativeness of a node can be computed based on the social relations and its behavior.

Scalability (Sc): For large IoT environments, scalability is an important factor, for example, a scalable TMS will play an important role if billions of IoT devices are communicating with each other with increased data, communications, and transactions. To be fully functional, TMS must be scalable to accommodate an increasing number of devices. With the increasing number of IoT devices, scalability must be taken into consideration during the development of a trust model. The real-time deployment of a TMS becomes very difficult when accommodating and storing the trust values from many devices. The distributed model requires some storage mechanisms and processes a large amount of information.

Energy (EN): Most IoT systems use battery-operated devices, and the energy of those devices is directly proportional to the amount of information those devices can process and communicate.

Reputation (RP): The reputation (RP) of a node is the probability of its behavior, based on other entity's observations about the actions of the nodes within a certain circumstance and time period.

Resilience (RE): IoT systems are associated with critical infrastructure where safety and resilience are paramount. Trust management must be resilient to trust-related attacks and malicious activities that target the trust propagation and calculation processes. Most of the schemes in the literature fail to achieve the resilience of trust management against notable misbehaving attacks, including on-off, non-trustworthy recommendations, and self-promoting attacks.

Table 3 highlights the most important challenges faced by traditional trust and reputation evaluation techniques in the IoT. This evaluation highlighted problems that weaken the TMS in the IoT scenario. IoT devices, for example, consume energy, generate dynamic data, suffer from communication delays, require context awareness, and require scalability. However, most of the TMS focused on energy efficiency and data dynamicity, but struggled with scalability and resilience. Furthermore, most of the connected devices are resource-constrained in terms of storage and energy. As per Table 3 the storage, energy efficiency, context awareness, and dynamicity of IoT devices are the most prominent features of TMS. However, most of the literature lacks the capabilities of energy efficiency, storage, and dynamicity.

## Artificial intelligence based models

With the advancement of AI in the IoT domain, current research shows the applicability of AI techniques to solve the problems of IoT devices, e.g., trust calculations, behaviour learning, security, and privacy [103, 104]. Wang et al. [105] developed a technique to calculate the trust values for the social network, the method utilised logistic regression to minimise loss function parameters. Upel et al. [106] proposed an intelligent trust calculation model that can generate an accurate trust assessment for prospective actors. The proposed scheme numerically calculates the individual trust features and classifies them to get final trust values. Researchers in [107, 108] reported various trust management schemes based on multiclass classification and reinforcement learning frameworks. Furthermore, a study in [109] used machine learning (ML) techniques to carry out system recommendations to observe nodes in machine-to-machine communication for finding malicious and trustworthy information. Similarly, a study in [110] investigate a Multiple Linear Regression Problem (MLRP) and employed a K-means clustering-based scheme to classify malicious nodes and achieved higher accuracy. Moreover, a study in [111], established a discriminative trust-based scheme for service provisioning in IoT. Chinnaswamy et al. [112] presented a trust management model that focuses on IoT safety concerns. However, in the following section, we further classified the AI-based model into two subcategories.

**Data mining based models**

The DM algorithms employed in IoT trust and reputation could be classified according to their simulations platform into (a) Direct Trust and (b) Assisting Trust as shown in the taxonomy of the IoT trust and reputation in Fig. 3. The IoT infrastructure and machine learning techniques enabled robust and intelligent applications by incorporating the capabilities to perceive, reason, decide, take action, learn and interact. The ML-based trust models are categorised into two main groups; 1) direct trust evaluation and 2) assisting trust evaluation. In the following, we provide a discussion of each category and its suitability for various IoT applications.

**Direct trust evaluation models**

Machine learning (ML): ML techniques are used in trust evaluation to calculate the trust value based on whether the calculated entity is trustworthy or not based on trust-related features. The following ML algorithm is based on learning granularity, which is divided into three types: binary, numeral, and hybrid rating. The following is a succinct overview of these algorithms.

Binary Ratings: For trust evaluation in binary ratings, the ML algorithm is categorised into two groups: trustworthy and non-trustworthy, and mostly the binary algorithm is used for these kinds of problems. The following listings explain these ML techniques.

Logistic Regression (LoR): A study in [113] investigated a trust model based on LoR that provides an accurate method in routing protocol for lossy network to classify the non-trustworthy nodes. The proposed model employed the LoR technique to identify the behavior of the nodes based on the combined trust values.

Numeral Ratings To leverage the numeral rating machine learning algorithm for the calculation of trust, the outcomes of evaluation are stated by discrete numbers. For example, the author of [114] employed the numbers 0, 1, and $-1$ to show the outcomes of trust evaluation, where 0 implies neutral, 1 implies trustworthy, and $-1$ shows non-trustworthy.

Naive Bayes Classifier: The authors of [115] discussed common problems and proposed a pervasive computing architecture based on a simple but effective trusted model. The proposed model combines some AI techniques to achieve the goal of human-like decision-making. As a result, the apriori algorithm is used to extract the behavioral patterns from data through the network interactions. Naive Bayes classifier is then used for final decision-making to communicate in terms of the possibility of the user's trustworthiness.

Hybrid Rating: A hybrid rating machine learning algorithm is appropriate to achieve both a binary and discrete number rating. The following terms demonstrate the technique used in hybrid rating.

K-Nearest Neighbor (KNN): A study in [111] introduces a discriminative model of trust (DMT), for service provisioning in SIoT. DMT uses object ratings. It is based on the data mining model, which compares the service query context and other raters' past query context. It considers social similarity, service importance, and the provider's energy. It converts the problem into three-dimensional space and uses weighted KNN to weigh the contribution of each of the past k experiences to estimate the trust value.

Decision Tree (DT): The author of [116] presented a trust management model for the social IoT to identify the behaviour of the nodes. The proposed algorithm employed the Jaccard Coefficient (JC) to calculate the similarities between the things and DT algorithm to classify the behaviour of the nodes. Similarly, a study in [117] suggested a trust-based model that employed a DT algorithm to identify malicious activities in the Internet of battlefield thing environment.

Random Forest (RF): RF is the combination of several distinct DT's. RF proves efficient for the high-dimensional, unbalanced, and estimating missing data. However, it is ineffective for low-dimension data. Janani et al. [113] developed an RF-based model that provided security to the IoT network and enhanced the trusted routing of the IoT scenario by identifying the sinkhole attack. The developed model was limited to a single attack and did not mention other privacy issues or attacks on IoT devices. Support Vector Machine (SVM): SVM classifies the objects based on the distance between the hyperplane and the support vector. In [112] investigated the trust authentication model based on ML. In the validation phase, gateways omit those nodes whose trust value is less than the basic threshold value. A support vector machine (SVM) was employed for the calculation of the trust threshold value over the gathered traffic data. The proposed model has been evaluated based on a few performance metrics like energy, delay, and CO. The researchers in [110] suggested two models, SVM and Gradient descent, to understand the reputation value of worker nodes and then cluster these nodes into trustworthy and non-trustworthy classes employing the K-means algorithm. For improvement in accuracy, they processed the learning model by improving the routing direction and calculating the values of trust for all nodes in the IoT environment.

Artificial Neural Networks (ANN): The researchers in [118] suggested an intelligent trust management system. the system utilised the ANN to make obtainable suggestions on the trustworthiness of each node and identifies the standard of an individual node, such as trustworthy or non-trustworthy.

Q-Learning: Aref et al. [119] proposed a hybrid trust model (HTM) that employs fuzzy logic and Q-learning for TM. The authors used the learning model for trust calculation and fuzzy logic for various aspects of trust aggregation.

**ML techniques for evaluation of assisting trust**

ML techniques play an auxiliary role in various tasks of trust evaluation. These techniques should be employed to execute the data for the calculation of trust.

K-Means Algorithm: In [120], the authors presented conditional packet manipulation attacks called targeted insider attacks. The proposed scheme maintains limited performance metrics of trust for each node, which show the possibility of initial attacks such as forwarding the packets with specific values. Moreover, the author uses a clustering and regression framework to calculate the node's trustworthiness and classify it into a trustworthy and non-trustworthy group.

DBSCAN Algorithm: A study in [121] investigated DBSCAN's task of selecting input parameters benefiting from correlation. This method has an advantage in execution, handling noise points, and high clustering speed. Furthermore, DBSCAN has high computation complexity like K-Means as the DBSCAN algorithm is executed using the entire dataset. However, DBSCAN is ineffective in resource-limited IoT with huge high-dimensional data. For example, when the density of data is uneven and the distance between clusters is far away then the condition of clustering should be unsatisfactory and inefficient.

Chameleon: Distributed chameleon hash parameter generation and trapdoor recovery methods can avoid the security problems faced by the centralised organisations and build an intelligent trust layer for IoT. The theoretical analysis and experiments show that the approach can effectively deal with malicious behaviours [122].

Even though these research accomplishments show some promising and elevated results by employing ML algorithms, they still lack the potential to be generic algorithm that can be commonly applied to any service domain without limiting it to specific infrastructures like MANET, WSN, and Underwater Acoustic Networks, etc. However, the limitations of these already established methods do not capture the subjectivity and context awareness of trust. Similarly, trust evidence, privacy protection, and computational overhead were not analysed in these techniques. Additionally, most of the literature considers quite limited performance metrics for the trust and reputation assessment processes, like delay, energy, computational overhead, and throughput. The following Table 4 presents a review and comparison of trust calculation based on artificial intelligence.

In Table 4, a comparative analysis of the existing schemes for trust management employing ML is carried out based on multiple parameters, such as subjectivity, robustness, context awareness, effectiveness, and computational overhead. We summarised the existing ML algorithms for TMS in various application scenarios, such as social networks, service-oriented systems, and the IoT domain. The algorithms used in such environments are supervised learning algorithms because such algorithms are widely used and easy to understand. In particular, many approaches use a set of classes to represent the results of trust calculations, which is insufficient to reflect trust uncertainty. Most schemes [115, 131, 132], were found to be effective in terms of accuracy. However, a few of the existing approaches [106, 119, 133], resist malicious activities in the evaluation of trust. Computational overhead is rare in some approaches like [115, 125, 127, 130] and does not pay special attention to evaluating the trust efficiency. Moreover, analysis of the computational overhead for an algorithmic scheme is valuable in determining the effectiveness of the system. The ML-based trust evaluation models should also consider the computation complexity and some other quality features of the system, such as accuracy, precision, and robustness to increase the applicability in resource constrained IoT devices.

**Table 4** Comparative Analysis of AI Trust-Based Models

| Ref. | Tech | TC | FS | CA | CO | PP | RB | Eff |
|---|---|---|---|---|---|---|---|---|
| [123] | KNN | Binary class | User info | ✗ | ✗ | ✗ | ✗ | Acc: 0.83 |
| [124] | RF | [0,1] | user features | ✗ | ✗ | ✗ | ✗ | Acc: 0.7 |
| [125] | SVM | [1, -1] | Mul. features | ✗ | ✗ | ✗ | ✗ | Acc: 0.81 |
| [126] | Reg | Binary classification | features | ✗ | ✗ | ✗ | ✗ | Acc: 0.73 |
| [127] | RBM | [1, -1] | User rating | ✗ | ✗ | ✗ | ✗ | Acc: 0.7-0.9 |
| [128] | RBM | NA | NA | ✗ | ✗ | ✗ | ✓ | ✗ |
| [114] | SVM | [1,0, -1] | NA | ✗ | ✗ | ✗ | ✗ | Acc: 0.97 |
| [119] | Fuzzy | Four categories | NA | ✓ | ✗ | ✗ | ✓ | ✗ |
| [115] | NB | [1,0, -1] | beh. features | ✗ | ✗ | ✗ | ✓ | AUC: 0.92 |
| [106] | SVM | [0,1] | credi bility info | ✗ | ✗ | ✗ | ✗ | Pre: 0.89 |
| [129] | SVM | [0,1] | features | ✗ | ✗ | ✗ | ✗ | Acc: 0.97 |
| [130] | DT | [1, -1] | NA | ✗ | ✗ | ✗ | ✗ | Acc: 0.90 |
| [131] | ANN | [1, -1] | Vehi cular | ✓ | ✗ | ✗ | ✓ | Precision: 0.92 |

### Deep learning based on IoT security and trust management

The application of deep learning (DL) to IoT schemes has become an interesting area of study. The key benefit of using DL instead of classical ML is its higher performance on large datasets. Many IoT devices generate a lot of data; as a result, DL approaches are appropriate for such systems. DL techniques should facilitate the deep linking of the IoT environment [132]. IoT-based devices and their applications can interact with one another automatically without human involvement using the deep-linked protocol. For example, in a smart home, the IoT devices automatically interact to form complete smart home automation. To learn data representations with various levels of abstraction, deep learning methods offer a computational architecture that integrates many processing layers. Modern applications have been greatly improved by DL approaches compared to previous ML algorithms. The design of DL is motivated by the operation and processes of the human brain. In this section, the Recurrent Neural Networks (RNNs), Autoencoders (AE), Sparse Autoencoder (SAE), Restricted Boltzmann Machines (RBMs), Deep Neural Network (DNN), and Long Short Term Memory (LSTM) will be discussed to show how these techniques are employed to identify trustworthy and non-trustworthy devices.

Deep Neural Networks (DNN): The study in [134] proposed an adaptive trust boundary for industrial IoT (IIoT) networks employing a deep neural network and a supervised learning algorithm. The developed scheme identified the types of attacks without any prior knowledge of their nature and did not require any manual attempts. Due to the large trust boundaries of IIoT networks, the developed system can adapt quickly to changes in attack models and can automatically update the recognition mechanism of the knowledge base, thereby preventing zero-day attacks on IIoT systems.

Recurrent Neural Network (RNN): RNN has a potential application in IoT devices, a study in [135] presented a reinforcement learning-based trust model using an Ad-hoc distance vector. The developed model involves a reinforcement learning manager for the detection of trustworthy and non-trustworthy nodes. The simulation parameters of the distance vector using a network simulator were provided to the designed RNN technique.

Long Short Term Memory (LSTM): LSTM is a DL algorithm that has gained the interest of the research community and has generated outstanding outcomes while employed in complex problems, such as the interpretation of languages, text generation, computerised captioning of pictures, and network security [136, 137]. Moreover, LSTM and multi-attribute rating techniques (SMART)were suggested in [138]. The proposed algorithm was employed to manage trust in IoT devices. However, the multi-attribute rating algorithm was employed to calculate the trust values, and LSTM was used to find out the trust threshold based on the behaviour changes. Another researcher [139] employed the Long Short-Term Memory (LSTM) approach to identify the history-based sequence of reputation values.

The drawbacks of existing schemes include changes in high memory utilisation to handle big data and the challenge to measure the uncertainty in untrusted behavior. Moreover, many techniques are based on the combined datasets as there are no actual datasets based on trust. The authors in [138, 139] examined more innovative solution schemes for privacy, and data integrity based on statistical and DL concepts. But such concepts mostly depend on data reputation and privacy and did not discuss the model validation.

### Open issues and challenges

This study identified open research challenges and future directions for evaluating trust and reputation management systems of IoT devices and classified them based on IoT trust and reputation functional requirements, as well as various AI-based trustworthy and malicious activities occurring in IoT systems.

1. One of the most challenging issues in IoT device security is the selection and composition of trust performance metrics. Furthermore, the user should use performance metrics to create trustworthy and honest recommendations. In the real world, some IoT devices belong to individual owners and are linked via the owner's social network, and several IoT trust-based applications are socially oriented, as shown in Table 3. Because relationships deserve importance in their recommendation due to social interest and interaction, social similarities, shared interests, and social metrics are used to aid and evaluate the trustworthiness of IoT devices. However, more research is required to enhance the trust computation in terms of accuracy and resilience against trust related attacks.

2. Much research work and investigation are needed to enhance and develop the mechanisms of the trust management system. There is still a research gap in the domain of trust modeling employing various performance metrics. These metrics should be used by assigning weights, either static or dynamic to all chosen trust metrics. In the existing work [33–35, 38, 42, 48, 49, 53, 66], IoT based TMS frequently employ single trust formation with weighted sums. As the multi-trust evaluation schemes consider various trust characteristics and metrics, therefore, developing multi-trust evaluation schemes will further enhance the trust establishment.

3. The limitations of low-power batteries and microcontrollers are the main issues with IoT devices. However, different machine learning techniques, such as DNN, require a large amount of computing power and resources. As a result, trust establishment using computationally expensive techniques must distribute the task across multiple processing nodes in order to save power and achieve performance efficiency.

4. Most of the researchers used a multi-model approach for trust management in IoT devices. For example, the authors of [138] combine the multi-attribute rating technique with the LSTM to calculate the values of trust and to identify the behaviour of the network nodes based on the trust threshold. Therefore, it is important to investigate further and use ensemble models to further enhance the performance in terms of accurate trust establishment.

5. ML and data mining techniques always consume a large amount of memory and additional resources during the processing of data generated by large IoT systems. If AI techniques are integrated with IoT devices, the system's computational complexity will increase. Therefore, it is necessary to design efficient machine learning techniques to minimise computational complexity, i.e., reduce the dimension of the data.

## Conclusion

With the advent of IoT, trust and reputation play a key role in the decision-making process of mitigating malicious devices, a challenging issue due to the scalable and dynamic nature of IoT. In this survey paper, we have established a taxonomy based on two main design perspectives for IoT trust and reputation, i.e., traditional trust management-based and artificial intelligence-based trust management. We have further classified traditional trust management based on two types (i.e., Decision and Evaluation models) and explained how researchers have employed the scheme and how users can assess the effectiveness of each scheme. By doing so, we have provided an in-depth understanding of the IoT's trust, reputation, safety conditions, and state-of-the-art schemes for ensuring the functional requirements as well as the technical procedures used by the systems. We have further discussed some of the important performance metrics utilised in trust evaluation schemes and identified current trends in trust and reputation management systems. Based on our analysis, we have emphasised that the existing approaches are still in the initial development stages and pointed out the need and directions for developing advanced trust management systems to mitigate the highlighted issues.

**Abbreviations**
| | |
|---|---|
| CA | Context awareness |
| OC | Online communication |
| CO | Communication overhead |
| DTMM | Dynamic Trust Management model |
| DM | Decision Models |
| CP | Computation Power |
| CA | Context Awareness |
| DT | Distributed Trust |
| TM | Trust Management |
| MN | Malicious Nodes |
| WSN | Wireless Sensor Network |
| FL | Federated Learning |
| EM | Evaluation Models |
| EE | Event Environment |
| TP | Trust Propagation |
| IDS | Intrusion detection system |
| TRM-IoT | Trust and Reputation model of IoT |
| BS | Base Station |
| DS | Dempster Shaper |
| TA | Trust aggregation |
| DSN | Distributed Sensor Network |
| QoS | Quality of Service |
| HTMM | Hybrid trust management model |
| DoS | Denial of service |
| DBN | Deep belief network |
| CTP | Centralised Trust Propagation |
| RNN | Recurrent neural network |
| MANET | Mobile ad-hoc network |
| FS | Features selection |
| LSTM | Long short-term memory |
| FST | Fuzzy Set Theory |
| DNN | Deep neural networks |
| ECR | Energy consumption ratio |
| MAGM | Multi-Attribute Graph Model |
| AIM | Artificial Intelligence Model |
| PP | Privacy protection |
| Ref. | References |
| Tech. | Techniques |
| FB | Feedback |
| EE | Event Environment |
| TC | Trust Characterisation |
| FS | Feature Selection |
| PP | Privacy Protection |
| Rb | Robustness |
| Sub | Subjectivity |
| Eff | Effectiveness |
| OC | Online Communication |
| CoP | Cooperativeness |
| SN | Social Network |
| Acc | Accuracy |
| RC | Recommendation |
| DY | Dynamicity |
| FTDAM | Fuzzy Tech. Dynamic Access Mechanism |
| CST | Computer Science and Theory |
| ✓ | Addressed |
| ✗ | Not addressed |

### Acknowledgements
The authors would like to thank Ulster University for supporting this work.

### Authors' contributions
Muhammad Aaqib wrote the survey paper. Aftab Ali, Liming Chen, and Omar Nibouche supervised the research and reviewed the paper. All authors gave final approval of the version to be published.

### Funding
This work received no specific grant from any funding agency.

### Availability of data and materials
No data were employed or evaluated during the current work.

## Declarations

### Ethics approval and consent to participate
Not Applicable

### Competing interests
The authors declare no competing interests.

Received: 26 October 2022   Accepted: 1 March 2023
Published online: 22 March 2023

**Publisher's Note**